\documentclass{JHEP3}
\usepackage{amssymb}

\newcommand{\mbf}{\mathbf}

\title{Duality invariance of non--anticommutative
 $\mathbf{ N={ 1\over 2}}$ supersymmetric
$\mathbf{ U(1)}$ gauge theory}

\author{\"{O}mer F. Dayi \\
 Physics Department, Faculty of Science and
Letters, Istanbul Technical University,
 TR--34469 Maslak--Istanbul,
Turkey, and\\
 Feza G\"{u}rsey Institute,
 P.O.Box 6, TR--34684
\c{C}engelk\"{o}y--Istanbul, Turkey \\
E-mail: \email{ dayi@itu.edu.tr,  dayi@gursey.gov.tr} }
\author{Lara T. Kelleyane\\
 Physics Department, Faculty of Science and
Letters, Istanbul Technical University,
 TR--34469 Maslak--Istanbul,
Turkey\\
E-mail: \email{kelleyanel@itu.edu.tr} }
\author{Kayhan \"{U}lker \\
 Feza G\"{u}rsey Institute,
 P.O.Box 6, TR--34684 \c{C}engelk\"{o}y--Istanbul, Turkey \\ E-mail:
\email{kulker@gursey.gov.tr} } 
\abstract{ A parent action is introduced to formulate 
(S--) dual of non--anticommutative $N={1\over 2}$ supersymmetric $U(1)$ 
gauge theory. Partition function for parent action in phase space is 
utilized to establish the equivalence of partition functions of the 
theories which this parent action produces. Thus, duality invariance of 
non--anticommutative $N={1\over 2}$ supersymmetric $U(1)$ 
gauge theory follows. The results which we obtained are valid at tree 
level or equivalently at the first order in the nonanticommutativity 
parameter $C_{\mu\nu}$.}

\keywords{S--duality, Supersymmetry, Noncommutativity}

\preprint{}

\begin{document}


\newcommand\be{\begin{equation}}
\newcommand\ee{\end{equation}}
\newcommand\fr{\frac}
\newcommand\tet{\theta}
\newcommand\bea{\begin{eqnarray}}
\newcommand\eea{\end{eqnarray}}
\newcommand\la{\lambda}
\newcommand\lb{\bar\lambda}
\newcommand\m{\mu}
\newcommand\n{\nu}
\newcommand\al{\alpha}
\newcommand\bet{\beta}
\newcommand\s{\sigma}
\newcommand\e{\epsilon}
\newcommand\ro{\rho}
\newcommand\ka{\kappa}

\newcommand\del{\partial}

\newcommand\Ra{\Rightarrow}
\newcommand\ra{\rightarrow}

\newcommand\ps{\slash\!\!\!\partial}
\newcommand\ds{\slash\!\!\!\nabla}

\newcommand\Ha{{\cal H}}
\newcommand\La{{\cal L}}
\newcommand\Ka{{\cal K}}
\newcommand\Da{{\cal D}}
\newcommand\Ma{{\cal M}}
\newcommand\Za{{\cal Z}}


\newcommand\Fbir{\Phi_1}
\newcommand\Fiki{\Phi_2}
\newcommand\Fd{\Phi_D}

\newcommand\fbir{\phi_1}
\newcommand\fiki{\phi_2}
\newcommand\fdbir{\phi_{D_1}}
\newcommand\fdiki{\phi_{D_2}}
\newcommand\fduc{\phi_{D_3}}
\newcommand\fddort{\phi_{D_4}}

\newcommand\Kbir{\chi_1}
\newcommand\Kiki{\chi_2}
\newcommand\Kd{\chi_D}

\newcommand\kbir{{\bar\chi}_1}
\newcommand\kiki{{\bar\chi}_2}
\newcommand\kd{{\bar\chi}_D}

\newcommand\vpbir{\varphi_1}
\newcommand\vpiki{\varphi_2}
\newcommand\vpuc{\varphi_3}
\newcommand\vpd{\varphi_D}

\newcommand\dbir{\Delta_1}
\newcommand\diki{\Delta_2}
\newcommand\dd{\Delta_D}


\newcommand\pmn{p_{\mu\nu}}

\newcommand\Pbir{P_1}
\newcommand\Piki{P_2}
\newcommand\Pd{P_D}

\newcommand\pdm{P_D^\mu}
\newcommand\pdo{P_D^0}
\newcommand\pdi{P_D^i}

\newcommand\pabir{\Pi_1^\alpha}
\newcommand\paiki{\Pi_2^\alpha}
\newcommand\pad{\Pi_D^\alpha}

\newcommand\pabbir{{\bar\Pi}_{1\dot\alpha}}
\newcommand\pabiki{{\bar\Pi}_{2\dot\alpha}}
\newcommand\pabd{{\bar\Pi}_{D\dot\alpha }}

\section{Introduction}

The formalism of superstring theory with pure spinors\cite{ber}
in a graviphoton background\cite{ov}
gives rise to a non-anticommutative superspace\cite{sei},\cite{nie} which 
was introduced also in other contexts \cite{mof}, \cite{klem}.
Moyal antibrackets (star products) are employed to interpose
non-anticommutativity between the coordinates. Thus,
instead of coordinates which are operators,
one deals with  the usual superspace variables.
Vector superfields taking values in  this deformed
superspace utilized to define a  non--anticommutative
supersymmetric Yang-Mills
gauge theory. However, due to a change of variables one deals
with the standard gauge transformations and component
fields\cite{sei}.
Deformation of 4 dimensional
$N=1$ superspace by making the chiral
fermionic coordinates $\theta_\al ,$ $\al =1,2,$  non--anticommuting
$$
\{\theta^\al ,\theta^\beta
\}=
C^{\al\beta} ,
$$
where $C^{\al\beta}$ are constant deformation parameters,
breaks half of the supersymmetry\cite{sei}.
In euclidean $\mathbb{R}^4$
chiral and antichiral fermions are
not related with
complex conjugation. The vector superfield of this deformed superspace
was employed to derive,
after a change of variables,
the $N=\frac{1}{2}$ supersymmetric
 Yang-Mills theory action\cite{sei}\footnote{For another approach see 
\cite{lerda}.}
\be
\label{is}
I_{1/2} = \fr{1}{g^2} \int d^4x Tr \Bigg\{ -\fr{1}{4} G^{\m\n} G_{\m\n}
- i \la \mathcal{D}\!\!\!\!/ \lb  + \fr{1}{2} D^2 - \fr{i}{2}  C^{\m\n} G_{\m\n}
(\lb\lb) + \frac{|C|^2}{8}(\lb\lb)^2
\Bigg\},
\ee
where $C^{\mu\nu}=C^{\al\beta}\epsilon_{\beta\gamma}
\sigma_{\;\al}^{\mu\nu\;\gamma}$ and $\mathcal{D}_\mu$ is the covariant derivative. Gauge transformations
possess the usual form.
$G_{\mu\nu}$ is the non--abelian field strength
related  to the gauge field $A_\mu .$
$\lambda\; ,\;\bar\lambda $ are independent fermionic fields
and $D$ is  auxiliary bosonic field.
Although we deal with euclidean $\mathbb{R}^4,$
we use Minkowski space  notation and
follow the conventions of \cite{wb}.
The surviving part of the $N=1$
supersymmetry acts on the standard component fields as
\bea
\delta\lambda &=& i\epsilon D +\sigma^{\mu\nu}
\epsilon(G_{\mu\nu}+\frac{i}{2}C_{\mu\nu}\bar\lambda\bar\lambda)\nonumber\\
\delta A_\mu &=& -i\bar\lambda{\bar\sigma}_\mu\epsilon \nonumber\\
\delta D &=& -\epsilon\sigma^\mu \mathcal{D}_\mu \bar\lambda \nonumber\\
\delta \bar\lambda &=& 0 \; \label{susy},
\eea
where $\epsilon$ is a constant Grassmann parameter.
The action (\ref{is}) can also be obtained by applying
the supersymmetry generator
$Q$ defined by $\delta =\epsilon Q, $
to the lower dimensional field monomial $Tr\la\la$
as
\be
I_{1/2}= \frac{1}{8g^2} Q^2 \int d^4 x Tr(\la\la),
\ee
up to total derivatives, similar to  the usual $ N=1$
super Yang-Mills theory\cite{kay}.

(S--) Duality transformations map strong coupling domains to weak 
coupling domains of gauge theories. Although duality invariance of pure 
$U(1)$ gauge theory can be shown, trivially, by rescaling its gauge 
fields\footnote{For $U(1)$ gauge theory rescaling $A\ra g^2 A_D$ 
results in the duality transformation $g^{-2}\int dA \wedge dA \ra g^2 \int 
dA_D \wedge dA_D$.}, it can also be studied  in terms of parent action 
formalism\cite{bus}. The latter approach permits to introduce a dual 
formulation of the  noncommutative $U(1)$ gauge theory\cite{grs}. 
Moreover, dual actions for supersymmetric $U(1)$ gauge  theory have 
already been derived utilizing a parent action when only bosonic
coordinates are noncommuting\cite{okb}. Actually, (S--) duality  is 
helpful for inverting computations performed in weak coupling domains to 
strong coupling domains, when partition functions of the ``original" and 
dual theories are equivalent, i.e. when there exists  duality symmetry. 
For noncommutative $U(1)$ gauge theory without supersymmetry, this 
equivalence was established  within the  hamiltonian formalism\cite{bo}.

We would like to investigate  duality properties of the $N={1 \over 2}$ 
supersymmetric non--anticommutative theory (\ref{is}) with $U(1)$ gauge 
group. We will define  dual theory by introducing a parent action which 
produces the original theory when ``dual" fields are eliminated by their 
equations of motion. Hamiltonian formalism of parent action is used to 
construct its partition function in phase space. We show that this 
partition function gives rise to either partition function of the original 
$N={1 \over 2}$ superysmmetric non--anticommutative $U(1)$ gauge theory
or partition function of its dual theory. Then, we conclude that the 
$N={1 \over 2}$ supersymmetric non--anticommutative $U(1)$ gauge theory is 
duality invariant. We do not consider loops, so that our results are valid 
at tree level which is equivalent to the first order approximation in 
$C_{\mu\nu}.$

In Section 2 we introduce a parent action. We show that it generates 
non-anticommutative $N={1\over 2}$ supersymmetric $ U(1)$ gauge theory. 
Then, we obtain obtain the dual theory resulting from this parent action. 
In  Section 3 we present hamiltonian formulation of the original and dual 
non-anticommutative $N={1\over 2}$ supersymmetric $ U(1)$ gauge theory. 
In Section 4, we first exhibit constrained hamiltonian structure arising 
from the parent action. Then, its path integral in the phase space is 
presented. By integrating over the appropriate variables we demonstrate 
the equivalence of partition functions of dual and original 
non-anticommutative $N={1\over 2}$ supersymmetric $ U(1)$ gauge theories. 
Lastly we comment on quantum corrections.


\section{Dual of non--anticommutative
 $\mathbf{ N={1\over 2}}$ supersymmetric  $\mathbf{ U(1)}$ gauge theory }

Parent action of supersymmetric $U(1)$ gauge theory by superfields was given in \cite{sw-p}. Once  written in terms of component fields it  was generalized to provide dual formulations of noncommutative supersymmetric $U(1)$ gauge theory when only bosonic coordinates of superspace
are mutually noncommuting\cite{okb}. By a similar approach we would like to introduce a parent action for the non--anticommutative $N={1\over 2}$ supersymmetric  $U(1)$ gauge theory obtained  from (\ref{is}). We propose the following parent action in terms of component fields
$\mathrm{X}=(F_{\mu\nu},\;\la_\al,\;\bar{\la}^{\dot\al},\;
\psi_\al,\;\bar{\psi}^{\dot\al},\;D_1,\;D_2)$
and $\mathrm{X}_D=(A_{D\mu},\;\la_{D\al},\;\bar{\la}^{\dot\al}_D,\;D_D )$,
\be \label{ip}
I_p = I_0 [\mathrm{X}] + I_l [\mathrm{X},\mathrm{X}_D]
\ee
where $I_0$ is suggested by (\ref{is})
\be \label{io}
I_0 = \fr{1}{g^2} \int d^4 x \Bigg\{
-\fr{1}{4}F^{\mu\nu}F_{\mu\nu} - \fr{i}{2}
\lambda\ps\bar\lambda - \fr{i}{2}
\bar\psi\bar{\ps}\psi + \fr{1}{4}D_1^2 + \fr{1}{4}D_2^2
-\fr{i}{4}C^{\mu\nu}F_{\mu\nu}(\bar\lambda\bar\lambda +
\bar\psi\bar\psi) \Bigg\}
\ee
and $I_l$ is defined as
\be
\label{il}
I_l = \int d^4 x\Bigg\{ \fr{1}{2}\epsilon^{\mu\nu\lambda\kappa}
F_{\mu\nu}\del_\lambda A_{D\kappa} + \fr{1}{2}\lambda\ps\bar\lambda_D +
\fr{1}{2}\lambda_D\ps\bar\lambda - \fr{1}{2}\bar\psi\bar{\ps}\lambda_D -
\fr{1}{2}\bar\lambda_D\bar{\ps}\psi +\fr{i}{2}D_D(D_1 - D_2) \Bigg\} .
\ee
Here  $F_{\mu\nu}$ are independent field variables which are not associated with any gauge field.

The equations of motion with respect to the ``dual" fields $\mathrm{X}_D$ are
\bea
 &&\e^{\m\n\la\ka}\del_\n  F_{\la\ka} = 0 \; , \label{eom}\\
&&\ps\bar\psi = \ps\lb \quad , \quad \bar{\ps}\psi = \bar{\ps}\la ,
\quad ,\quad D_1 = D_2 = D \; . \label {eqs1}
\eea
One solves
(\ref{eom})  by setting $F_{\mu\nu} =\del_\mu A_\nu - \del_\nu A_\mu$ which is the
field strength of the gauge field $A_\mu$. When one  plugs this and the solutions of the other
equations of motion (\ref{eqs1}) in terms of $\lambda ,\ \bar{\lambda},\ D ,$
into the parent action (\ref{ip}),  the non--anticommuting $N={1\over 2}$ supersymmetric
$U(1)$ gauge theory action follows:
\be
I = \fr{1}{g^2} \int d^4x \Bigg\{ -\fr{1}{4} (\del_\mu A_\nu - \del_\nu A_\mu)^2  - i \la \ps \lb  + \fr{1}{2} D^2 - \fr{i}{2}  C^{\m\n} (\del_\mu A_\nu - \del_\nu A_\mu) \lb\lb
\Bigg\} .\label{i}
\ee
Since we deal with $U(1)$ gauge group, the term quadratic in the deformation parameter, 
$\frac{|C|^2}{8}(\lb\lb)^2,$ of the action (\ref{is}) vanishes.

On the other hand, the equations of motion with respect to the fields  $\mathrm{X}$ are 
\bea
 &&\fr{1}{2g^2} F^{\m\n} + \fr{i}{4g^2} C^{\m\n} ( \lb\lb +
\bar\psi\bar\psi ) - \fr{1}{2} \e^{\m\n\la\ka} \del_\la A_{D_\ka} = 0 \;
,\nonumber\\
 &&\ps\lb + ig^2\ps\bar\lambda_D =0\quad , \quad \ps\bar\psi - ig^2\ps\bar\lambda_D =0\;
,\nonumber\\
&& \bar{\ps}\la +C^{\m\n} F_{\m\n} \lb + ig^2 \bar{\ps}\la_D = 0 \quad , \quad
\bar{\ps}\psi + C^{\m\n} F_{\m\n} \bar\psi -i g^2 \bar{\ps}\la_D = 0 \; ,
\nonumber\\
&& D_1 + ig^2 D_D=0 \quad , \quad  D_2 - ig^2 D_D =0 \; \label{eomd}
\eea
where $\;F_{D\mu\nu}=\del_\mu A_{D\nu} - \del_\nu A_{D\mu}$. We solve the equations of 
motion (\ref{eomd}) for $\mathrm{X}$ fields in terms of $\mathrm{X}_D$ and substitute 
them in the parent action (\ref{ip}) to obtain the dual non--anticommutative  $N={1\over 2}$ supersymmetric $U(1)$ gauge theory action :
\be \label{id}
I_{D} = g^2 \int d^4 x \Bigg\{ -\fr{1}{4}F_D^{\m\n}F_{D\m\n}-
 i\la_D
\ps\lb_D + \fr{1}{2}D_D^2 + \fr{i}{4}g^2\e^{\m\n\la\ka} C_{\m\n}
F_{D\la\ka} \lb_D\lb_D \Bigg\}.
\ee

One can observe that the non--anticommutative  $N={1\over 2}$ supersymmetric $U(1)$ gauge theory action (\ref{i}) and its dual (\ref{id}) possess the same form and
\bea
g & \rightarrow & \fr{1}{g} \nonumber\\
C^{\mu\nu} & \rightarrow &C^{\mu\nu}_D =
-\fr{1}{2} g^2\e^{\m\n\la\ka}C_{\la\ka} =ig^2 C^{\mu\nu} \label{cd}
\eea
is the duality transformation.

\section{Hamiltonian formulations of non--anticommutative $\mathbf{ N={1\over 2}}$ 
\mbox{supersymmetric} $\mathbf{ U(1)}$ gauge theory and its dual}

To acquire hamiltonian formulation of the non--anticommutative $ N={1\over 2}$ 
supersymmetric $U(1)$ gauge theory (\ref{i}), let us introduce the canonical momenta 
$\;(P^{\mu},\; \Pi^\al,\;\bar{\Pi}_{\dot\al},\;P)$ corresponding to 
$\;(A_\mu,\;\la_\al,\;\bar{\la}^{\dot\al},\;D).$ Canonical momenta associated with $A_i,$ $i=1,2,3;$ are
\be
P^i = -\frac{1}{g^2}(\del^0 A^i - \del^i A^0) - \frac{i}{g^2}
C^{0i}\la \la .
\ee
However, definitions of the other momenta lead to the weakly vanishing primary constraints\footnote{We use left derivatives with respect to  anticommuting variables throughout this work.},
\bea
\phi_1 \equiv P^0 \approx 0 \quad &,&\quad  \Phi_1 \equiv P \approx 0 \nonumber\\
\chi^\al \equiv \Pi^\al \approx 0 \quad &,&\bar{\chi}_{\dot\al} \equiv \bar{\Pi}_{\dot\al} -  \frac{i}{g^2}\la^\al \s^0_{\al\dot\al} \approx 0 \; \label{pc}.
\eea
Canonical hamiltonian associated with the action (\ref{i}) is derived to be
\bea
\Ha_c
&=& \fr{g^2}{2}P^2_i
+\frac{1}{g^2}\lbrace \fr{1}{4} (\del_i A_j - \del_j A_i)^2 +i\la\ds{\lb} - \fr{1}{2}D^2 + \fr{i}{2}C^{ij}(\del_i A_j - \del_j A_i)\lb\lb\rbrace\nonumber\\
&&- A_0\del_i P^i -i C_{0i}P^i\lb\lb \label{hc}
\eea
where $\ds=\sigma^i\del_i.$

Let the primary constraints (\ref{pc}) be collectively denoted as $\Theta^a$. Then the extended hamiltonian is given by
\be
\label{exth}
\Ha_E = \Ha_c + l_a\Theta^a\; ,
\ee
where $l_a$ are Lagrange multipliers. Consistency of the primary constraints (\ref{pc}) with the equations of motion following from (\ref{exth}):
$$
\dot\Theta^a = \{ \Ha_E,\Theta^a \}  \approx 0
$$
gives rise to the secondary constraints
\be \label{sc}
\phi_2 \equiv \del_i P_i \approx 0 \quad ,\quad \Phi_2 \equiv D \approx 0 \;  .
\ee
There are no other constraints arising from these secondary constraints. One can show that $\phi_1 ,\;\phi_2$ are first class and $\Phi_1,\; \Phi_2, \; \chi^\al ,\; \bar{\chi}_{\dot\al}$ are second class constraints.

Hamiltonian structure  of the dual theory (\ref{id}) is similar to 
(\ref{pc})--(\ref{sc}). 
Indeed, canonical  hamiltonian associated with the dual action (\ref{id}) can easily be read from (\ref{hc}) as
\bea
\Ha_{Dc} &=& \fr{1}{2g^2}P_D^i P_{Di} +
g^2\lbrace\fr{1}{4}F^{ij}_D F_{Dij} +i\la_D\ds{\lb}_D - \fr{1}{2}D_D ^2 + \fr{i}{2}C_D^{ij}F_{ij}\lb_D\lb_D\rbrace
\nonumber\\
&& - A_{D0}\del_i P_D ^i -i C_{D0i}P_D ^i\lb_D\lb_D
\label{hdc}.
\eea
Moreover, there are the hamiltonian constraints which can be obtained from  (\ref{pc}) 
and (\ref{sc}) by  replacing $(P^0,\; \Pi^\al ,\;\bar{\Pi}_{\dot\al},\;P,
\; A_i,\;\la_\al,\;D)$ with $\;(P_D^0,\; \Pi_D^\al ,\;\bar{\Pi}_{D\dot\al},\;P_D,
\;A_{Di},\;g^4\la_{D\al},\;D_D)$. 


\section{Equivalence of partition functions for non--anticommutative
 $\mathbf{ N={1\over 2}}$ \\ 
 \mbox{supersymmetric} $\mathbf{ U(1)}$ theory and its dual}

Partition function for the parent action (\ref{ip}) is expected 
to produce partition
functions of the actions (\ref{i}) and (\ref{id}).
In (\ref{ip}) there are some terms cubic in fields. Thus, it would be apposite to discuss
its partition function in phase space, where integrations would be simplified due to
hamiltonian constraints. 
To achieve hamiltonian formulation let us introduce the set of
canonical momenta 
$\;(P^{\mu\nu},\; \Pi^\al_1,\;\bar{\Pi}_{1\dot\al}, \;\Pi^\al_2,
\;\bar{\Pi}_{2\dot\al},\;P_1 ,\; P_2) $ 
and $\;(P^{\mu}_D,\;\Pi^\al_D,\;\bar{\Pi}_{D\dot\al},\;P_D)$ 
corresponding
to the fields $\;(F_{\mu\nu},\;\la_\al,\;\bar{\la}^{\dot\al},\;
\psi_\al,\;\bar{\psi}^{\dot\al},\;D_1,\;D_2)$ and 
to the dual variables
$\;(A_{D\mu},\;\la_{D\al},\; \bar{\la}^{\dot\al}_D,\;D_D ).$ 
Each of the canonical
momenta resulting from the parent action (\ref{ip}) gives rise to a primary constraint,
which we collectively denote them as $\{\Theta^a\}$ : 
\bea \fbir^{0i} \equiv P^{0i}
\approx 0 ,\quad &&\quad \fiki^{ij} \equiv P^{ij} \approx 0 \; ,\nonumber\\ \Kbir^\al
\equiv \pabir \approx 0 ,\quad & &\quad {\bar\chi}_{1\dot\al} \equiv \pabbir -
\fr{i}{2g^2} \la^\al \s^0_{\al\dot\al} + \fr{1}{2} \la^\al_D \s^0_{\al\dot\al} \approx 0,
\nonumber\\ \Kiki^\al \equiv \paiki - \fr{i}{2g^2} \bar\psi_{\dot\al}
\bar\s^{0\dot\al\al} - \fr{1}{2} \lb_{D\dot\al} \bar\s^{0\dot\al\al} \approx 0, \quad &&
\quad \chi_{2\dot\al} \equiv \pabiki \approx 0, \nonumber\\ \Fbir \equiv \Pbir \approx 0,
\quad &&\quad \Fiki \equiv \Piki \approx 0 , \nonumber\\ \fdbir \equiv \pdo \approx 0
,\quad &&\quad \fdiki^i \equiv \pdi - \fr{1}{2} \e^{ijk} F_{jk} \approx 0 , \nonumber\\
\Kd^\al \equiv \pad - \fr{1}{2} \bar\psi_{\dot\al} \bar\s^{0\dot\al\al} \approx 0, \quad
&&\quad \chi_{D\dot\al} \equiv \pabd + \fr{1}{2} \la^{\al} \s^0_{\al\dot\al} \approx 0 ,
\nonumber\\ \Fd \equiv \Pd \approx 0. \quad && \label{pcp} 
\eea 
Canonical hamiltonian
associated with the parent action (\ref{ip}) is then found to be \bea \Ha_{p}
 &=& \fr{1}{g^2} \Big[ \fr{1}{4} F^2_{\m\n} + \fr{i}{2}
\la\ds\lb + \fr{i}{2}\bar\psi\bar{\ds}\psi - \fr{1}{4} (D_1^2 + D^2_2) + \fr{i}{4}C^{\m\n} F_{\m\n} (\lb\lb + \bar\psi\bar\psi) \Big] 
\nonumber\\
&& - \e^{ijk} F_{0i} \del_j A_{D\ka} + \fr{1}{2} \e^{ijk}F_{ij}\del_\ka A_{D0} 
- \fr{1}{2} \la\ds{\lb}_D -\fr{1}{2}\la_D \ds\lb + \fr{1}{2}\bar\psi\bar{\ds}\la_D 
\nonumber\\
&& + \fr{1}{2} {\lb}_D\bar{\ds}\psi - \fr{i}{2} D_D (D_1 - D_2). \label{hcp}
\eea
Extended hamiltonian is obtained by adding the primary constraints $\Theta^a$ with the help of Lagrange 
multipliers $l_a,$ to the canonical hamiltonian (\ref{hcp}):
\be
\Ha_E = \Ha_p + l_a\Theta^a
\ee
Consistency of the primary  constraints with the equations of motion:
$$
\dot\Theta^a =\{\Ha_E ,\Theta^a \} \approx 0
$$
gives rise to the secondary constraints
\bea
\dbir  &\equiv&  \{ \Ha_p, \Pbir \} = -\fr{1}{2g^2}D_1 - \fr{i}{2} D_D  \approx 0 \; ,\nonumber\\
\diki  &\equiv& \{ \Ha_p, \Piki \} = -\fr{1}{2g^2}D_2 +\fr{i}{2} D_D  \approx 0 \; ,
\nonumber\\
\dd  &\equiv&   \{ \Ha_p, \Pd \} = \fr{i}{2} ( D_1 -D_2 ) \approx 0 \; ,
\nonumber\\
\vpd &\equiv& \{ \Ha_p, \pdo \} = \fr{1}{2} \e^{ijk} \del_k F_{ij}  \approx 0 \; ,
\nonumber\\
\vpbir^{0i}  &\equiv & \{ \Ha_p,P_{0i} \} =
F^{0i} - g^2\e^{ijk} \del_j A_{Dk} + \fr{ig^2}{2} C^{0i} (\lb\lb +
\bar\psi\bar\psi)  \approx 0  \; .
\eea

In path integrals first and second class constraints are treated on different grounds.
Thus, let us first identify the first class constraints:  $\phi_{D1}$ is obviously first class. Moreover, we observe that the linear combination
\be
\label{xi}
\fduc \equiv \del_i \phi_{D_2}^i +\varphi_D=\del_i \pdi \approx 0,
\ee
is also a first class constraint. There are no other first class constraints. However, the constraints $\phi_{D_2}^i$ contain second class constraints 
which we should separate out. 
This is due to the fact that a vector can be completely described 
by giving its divergence and rotation (up to a boundary condition). (\ref{xi}) is derived taking divergence of $\phi_{D_2}^i,$ so that, there are still
two linearly independent second class constraints following
from the curl of $\phi_{D_2}^i:$
\be
\label{klo}
\fddort^n \equiv
K_i^n\phi^i _{D_2}= \Ka^{ni}\e_{ijk}\del^j \fdiki^k \approx 0,
\ee
where $n=1,2.$ $\Ka^n_i$ are some constants whose explicit forms are not needed for the purposes of this 
work. Although all of them are second class, we would like to separate $\vpbir^{0i} $ in a similar manner:
\bea
\vpiki & \equiv &\del_i \vpbir^{0i}= -\del_i F^{0i} 
- \fr{i}{2} C^{0i} \del_i(\lb\lb + \bar\psi\bar\psi) \approx 0, \\
\vpuc^n &\equiv & L_i^n\varphi_1^{0i}  = \La^{ni} \e_{ijk} \del^j\vpbir^{0k} \approx 0. \label{llo}
 \eea
where $\La^{nj}$ are some constants. The reason of preferring this set of constraints 
will be clear when we perform the path integrals, though  explicit forms of $\La^n_i$ play no role in 
our calculations.

In phase space, partition function can be written as\cite{fra},\cite{sen}
\bea \label{za}
\Za &=& \int \prod_i\Da \mathrm{Y}_i\;\Da P_{\mathrm{Y}_i} \;  \Ma \;
e^{i\int d^3 x
( \dot\mathrm{Y}_i P_{\mathrm{Y}_i} - \Ha_p)} \\
\Ma &=& N {\rm det}(\partial_i^2)
\delta(\mbf{\del} \cdot \mbf{P}_D)
\delta(\mbf{\del} \cdot \mbf{A}_D)
\delta (P_{D0})
\delta (A_{D0})
\textrm{sdet} \; M \; \prod_{z} \delta (S_z), \label{z0}
\eea
where $\mathrm{Y}_i$ and $P_{\mathrm{Y}_i}$ embrace  all of the
fields and their momenta.
$S_z$ denotes all second class constraints:
$ S_z\equiv (\fbir$, $\fiki$, $\Fbir$,
$\Fiki$, $\fddort$, $\Fd$, $\vpiki$,
$\vpuc$, $\dbir$, $\diki$, $\vpd$, $\dd$,
$\Kbir$, ${\bar\chi}_1$, $\Kiki$, $\kiki$, $\Kd$, $\kd ).$
We adopted the gauge fixing (auxiliary) conditions
\bea
A_{D_0} &=& 0 \; ,\nonumber\\
\del_i A_{D_i} &=& 0 ,
\eea
for the first class constraints $\fdbir $ and $\fduc $. $N$ is a normalization constant.
The matrix of the generalized Poisson brackets of the second class constraints
$M =\{S_z,S_{z^\prime}\}$ can be written in the form
\be
M = \left[\begin{array}{cccc}
      A & B\\
      C & D\end{array}\right] ,
\ee
so that,
its  superdeterminant  is given by
\be
\label{sd}
\textrm{sdet} M = ( \det D )^{-1} \det ( A - BD^{-1} C ).
\ee
Calculations of $B,C$ and $D$ can be shown to yield
$$( BD^{-1}C) = 0 .$$
Therefore, (\ref{sd}) is simplified as 
\be
\textrm{sdet} M = \fr{\det A}{\det D}.
\ee
Contribution of fermionic constraints is
\be
\det D^{-1} =  -(4 \det{g^2})^2.
\ee
Here, $\det{g^2}$, which arise because we deal with constraints of a field theory, should 
appropriately be regularized. Contribution of the bosonic constraints has already been calculated in \cite{bo}:
\be
\label{dsc}
\det A
= \det \left( \epsilon_{ijk}\del^iK_1^jK_2^k  \right)
\det \left( \epsilon_{ijk}\del^iL_1^jL_2^k  \right).
\ee
The linear operators $K_i^n$ and $L_i^n$ are defined in (\ref{klo}) and (\ref{llo}). These determinants which are  multiplication of three linear operators should be interpreted as  multiplication of their eigenvalues.

In (\ref{za}) the integrals over all of the fermionic momenta and $P_{\mu\nu}$ can be 
easily performed utilizing the related delta functions, to get
\bea
\Za &=& \int \Da F^{\m\n} \; \Da \la \;\Da \psi \; \Da \lb \; \Da \bar\psi \;
\Da D_1 \; \Da P_1 \; \Da D_2 \; \Da P_2\; \Da A_{D\m} \; \Da \la_D \; \Da \lb_D
\; \Da P_{D\m}  \; \Da D_D  \; \Da P_D
\nonumber\\
&& \tilde{\Ma}\;  \textrm{exp} \Bigg\{
i\int d^3 x \Bigg[
P_1\dot{D_1} +P_2\dot{D_2} + P_D^0 \dot{A}_{D0} + P_D^i \dot{A}_{Di} + P_D \dot{D}_D
-\fr{1}{4g^2}F^{0i} F_{0i}
\nonumber\\
&&-\fr{1}{4g^2}F^{ij} F_{ij}- \fr{i}{2g^2}\lambda\ps\bar\lambda -
\fr{i}{2g^2}\bar\psi\bar{\ps}\psi  + \fr{1}{4g^2} (D_1^2 + D_2^2)
-\fr{i}{2g^2}C^{0i} F_{0i}(\bar\lambda\bar\lambda + \bar\psi\bar\psi)
\nonumber\\
&&-\fr{i}{4g^2}C^{ij} F_{ij}(\bar\lambda\bar\lambda + \bar\psi\bar\psi)
+ \epsilon^{ijk} F_{0i}\del_j A_{Dk}
- \fr{1}{2}\epsilon^{ijk} F_{ij}\del_k A_{D0}
\nonumber\\
&&+ \fr{1}{2}\lambda\ps\bar \lambda_D
+ \fr{1}{2}\lambda_D\ps\bar\lambda - \fr{1}{2}\bar\psi\bar{\ps}\lambda_D
- \fr{1}{2}\bar \lambda_D\bar{\ps}\psi +\fr{i}{2}D_D(D_1 - D_2)
\Bigg] \Bigg\}  \label{z1} .
\eea
Here, $\tilde{\Ma}$ is the same with $\Ma$ except the delta functions which we utilized above. We first would like to integrate over the fields which do not carry the  label $``D"$ : $\Pbir$, $\Piki$ integrals are trivially performed and by integrating over $D_1$ and $D_2$ we get a factor of $\det{g^2}$ and $\delta(D_D)$. Integrations over $\psi$ and $\la$ yield  $({\det}\ps / \det g^2)^2 \delta(i\bar\psi + g^2 \lb_D) \delta(i \lb - g^2\lb_D)$. Thus, we
replace $\bar\psi$ with $ig^2 \lb_D$ and $\lb$ with $-ig^2 \lb_D$ after integrating over $\bar\psi$ and $\lb$. Integrations over $F^{\m\n}$ yield substitution of $F^{0i}$ with 
$g^2 \e^{ijk} \del_j A_{Dk} + \fr{i}{2} g^4 C^{0i} \lb_D\lb_D$, $F^{ij}$ with
$ \e^{ijk} P_{Dk}$ and cancellation of the determinant (\ref{dsc}). Moreover, we integrate over $A_D^0,\ P_D^0 $ and choose the normalization constant $N$ such that we get
\bea
\Za &=&
\int \; \Da A_{Di} \;\Da \lb_D \; \Da P_{Di} \; \Da D_D  \; \Da P_D \;
(\det{g^2})\det{\del_i^2})({\det}\ps )^2\; \delta(D_D) \delta (P_D)\;
\nonumber\\
&& \delta(\mbf{\del} \cdot \mbf{P}_D)
\delta(\mbf{\del} \cdot \mbf{A}_D)
\;\;\textrm{exp} \Bigg\{ i \int d^3 x
\Bigg[ P_D^i \dot A_{Di} + P_D \dot D_D - \fr{1}{2g^2} P_{Di} P^i_D - i C_{D}^{0i} P_{Di} \lb_D \lb_D \nonumber\\
&& - \fr{g^2}{4} F_D^{ij} F_{Dij}  - \fr{i}{2} g^2 C_D ^{ij}
F_{Dij} \lb_D \lb_D - ig^2 \la_D \ps \lb_D + \fr{g^2}{2} D_D^2\Bigg] \Bigg\}.
\label{z2}
\eea
In the exponent we distinguish the first order lagrangian of the dual theory
(\ref{hdc}) where  $\Pi^\al_D $ and $\bar{\Pi}_{D\dot\al}$ are eliminated from
the path integral by 
performing their integrations.

Now, in (\ref{z1}) let us integrate over the fields carrying the label $``D"$: $\Pd$ 
integral is trivial. Integration over  $D_D$ contributes as 
$(\det g^2)\delta(D_1+D_2)\delta (D_1 -D_2)$.  Integrations of the fermionic 
variables $\la_D$ and $\lb_D$ lead to 
$\delta(-\ps\bar\psi + \bar{\ps}\psi) \delta(\bar{\ps}\la - \bar{\ps}\psi)$. 
Due to the constraint $\varphi_D=0$ we set
\be\label{fij}
F_{ij}=\del_iA_j -\del_jA_i .
\ee
However, this replacement does not diminish the relevant number of physical phase space 
variables as it should be the case if the second class constraint
$\varphi_D$ has been taken properly into account. 
Therefore, we adopt the change of variables (\ref{fij}) with the replacement \cite{bo} 
\be
\label{nor}
\Da F_{ij} \delta (\e^{klm}\del_kF_{lm})
\delta (K_n^i( P_{Di}+\frac{1}{2}\e_{ijk}F^{jk}) )
 \to \det (\del^2)
\Da A_i \delta (\del_jA^j)
\delta \left(K_n^i( P_{Di}+\e_{ijk}\del^jA^k) \right).
\ee
Expressing $A_{Di}$ and $P_{Di}$ in terms of the fields $(A_i,\ F_{0i})$
by making use of the delta functions $\delta (K_i^n\phi_D^i) \delta (L_i^n\phi_1^{0i})
\delta(\mbf{\del} \cdot \mbf{P}_D) \delta(\mbf{\del} \cdot \mbf{A}_D) $
contributes to the measure with
\[
\left[ (\det g^2)^2 \det (\del^2)
 \det \left( \e_{ijk}\del^iK_1^jK_2^k  \right)
\det \left( \e_{ijk}\del^iL_1^jL_2^k  \right) \right]^{-1}.
\]
Hence, integrations over $A_{Di}$ and $P_{Di}$ in (\ref{z1}) can be performed to obtain
\bea
\Za &=& \int  \; \Da A_i \; \Da F_{0i} \Da \la \; \Da \lb \; \Da \psi \; \Da \bar\psi \;
\Da D_1 \; \Da \Pbir \; \Da D_2 \; \Da \Piki\;
(\det g^2) \det (\del_i^2)\;
\delta(\mbf{\del} \cdot \mbf{A})
\nonumber\\
&& \delta (D_1+D_2)\delta (D_1 - D_2) \; \delta (-\ps\bar\psi + \ps\lb) \;
\delta
(\bar{\ps}\la - \bar{\ps}\psi) \; \delta\left( \del_i F^{0i} +
\fr{i}{2}\del_iC^{0i} (\lb\lb + \bar\psi\bar\psi)\right) \;
\nonumber\\
&&   \textrm{exp} \Bigg\{ i \int d^3 x \Bigg[
\fr{1}{g^2}\left(  F^{0i} + \fr{i}{4} C^{0i}(\lb\lb + \bar\psi
\bar\psi)\right) \dot{A}_i + \dot{D}_1 \Pbir + \dot{D}_2 \Piki -
\fr{1}{2g^2}F^{0i} F_{0i}  
\nonumber\\
&& - \fr{1}{4g^2} (\del_i A_j - \del_j A_i)^2 - \fr{i}{2g^2} \lambda\ps\bar\lambda - \fr{i}{2g^2} \bar\psi\bar{\ps}\psi + \fr{1}{4g^2} (D_1^2 +
D_2^2) 
\nonumber\\
&&- \fr{i}{2g^2} C^{0i} F_{0i} (\lb\lb + \bar\psi\bar\psi) - \fr{i}{4g^2} C^{ij} (\del_i A_j - \del_j A_i) (\lb\lb + \bar\psi\bar\psi)
\Bigg] \Bigg\}  \label{z3}
\eea
Integrating over $D_2$, $\Piki$, $\psi$, $\bar\psi$ and renaming $D_1 =D$ and $P_1 =P$ yield
\bea
\Za &=& \int \; \Da A_i \; \Da F_{0i} \Da \la \; \Da \lb \; \Da D \; \Da P \;
(\det{g^2}) (\det{\del^2_i})(\det{\ps})^2 \delta (P)\delta (D)
\delta(\mbf{\del} \cdot \mbf{A})\nonumber\\
&&
\delta\left(\del_i F^{0i} + i\del_iC^{0i} \lb\lb
\right)  \textrm{exp}
\Bigg\{ i \int d^3 x \Bigg[ \fr{1}{g^2}\left( F^{0i} 
+iC^{0i} \lb\lb \right) \dot{A}_i + \dot{D} P
 - \fr{1}{2g^2}F^{0i} F_{0i} 
\nonumber\\
&& 
- \fr{1}{4g^2}(\del_i A_j - \del_j A_i)^2
 - \fr{1}{g^2} \lambda\ps\bar\lambda + \fr{1}{2g^2} D^2 
- \fr{i}{g^2} C^{ij} (\del_i A_j - \del_j A_i) \lb\lb
\Bigg] \Bigg\} . \label{zans}
\eea
In  terms of the change of variables
\bea
g^2 P^i &=& F^{0i} +  C^{0i} \lb\lb , \nonumber\\
\Da F^{0i} &=& (\det{g^2}) \Da P^i ,
\eea
we write the partition function (\ref{zans})
as
\bea
\Za &=& \int \; \Da A_i \; \Da P^i \; \Da \la \; \Da \lb \; \Da D \; \Da P\;
(\det{g^2}) \; (\det{\del^2 _i}) (\det{\ps})^2\; \delta (D) \delta (P)
\delta(\mbf{\del} \cdot \mbf{P})
\delta(\mbf{\del} \cdot \mbf{A})
 \nonumber\\
&& \textrm{exp} \Bigg\{ i \int d^3 x \Bigg[ P^i
\dot{A}_i + \dot{D}_1 \Pbir  - \fr{g^2}{2}(P_i)^2 - i C^{0i} P_i
\lb\lb -  \fr{1}{4g^2} (\del_i A_j - \del_j A_i)^2 \nonumber\\
&&  - \fr{i}{g^2} \lambda\ps\bar\lambda + \fr{1}{2g^2} D^2 - \fr{i}{2g^2} C^{ij} (\del_i A_j - \del_j A_i) \lb\lb \Bigg] \Bigg\}.  \label{z4}
\eea
In the exponent we recognize the first order lagrangian of the original theory
(\ref{hc}) after integrations over  $\Pi^\al_1,\;\bar{\Pi}_{1\dot\al}, 
\;\Pi^\al_2$ and  $\bar{\Pi}_{2\dot\al}$ are performed in its path integral.

Let us adopt the normalization to write  partition function of non--anticommutative
 $N={1\over 2}$ supersymmetric $ U(1)$ gauge theory as
\bea
Z_{NA} &=&
\int \;  \Da A_i \; \Da P_i \; \Da \la \; \Da \lb \; \Da D \; \Da P \;
\delta (D) \delta (P)\;
\delta(\mbf{\del} \cdot \mbf{P})
\delta(\mbf{\del} \cdot \mbf{A})
\nonumber\\
&&  \textrm{exp} \Bigg\{ \fr{i}{\hbar} \int d^3 x \Bigg[ P^i
\dot{A}_i + \dot{D} P  - \fr{g^2}{2}(P_i)^2 - i C^{0i} P_i
\lb\lb
\nonumber\\
&& -  \fr{1}{4g^2} (\del_i A_j - \del_j A_i)^2
- \fr{i}{g^2} \lambda\ps\bar\lambda + \fr{1}{2g^2} D^2
- \fr{i}{2g^2} C^{ij} (\del_i A_j - \del_j A_i)
\lb\lb   \Bigg] \Bigg\}.  \label{zn}
\eea
Therefore, by the applying the transformation (\ref{cd}), partition function of its dual
can be obtained  as
\bea
Z_{NAD} &=&
\int  \;  \Da A_i \; \Da P_i \; \Da \la \; \Da \lb  \; \Da D \; \Da P \;
\delta (D) \delta (P)
\delta(\mbf{\del} \cdot \mbf{P})
\delta(\mbf{\del} \cdot \mbf{A})
 \nonumber\\
&& \textrm{exp} \Bigg\{ \fr{i}{\hbar} \int d^3 x \Bigg[ P^i
\dot{A}_i + \dot{D} P  - \fr{1}{2g^2}(P_i)^2 - \fr{ig^2}{2} C_D ^{0i} P_i
\lb\lb
\nonumber\\
&&-  \fr{4g^2}{4}
 (\del_i A_j - \del_j A_i)^2
- i g^2 \lambda\ps\bar\lambda
+ \fr{g^2}{2} D^2   - \fr{ig^4}{2} C_D ^{ij}
 (\del_i A_j - \del_j A_i)
\lb\lb  \Bigg] \Bigg\}.  \label{znd}
\eea
Here, we omitted the label $``D"$ of the dual fields.

Comparing (\ref{z2}) and (\ref{z4}) which are obtained from the
 partition function for parent action (\ref{za}), by integrating 
different set of fields, one  concludes that the partition 
functions of non--anticommutative  $N={1\over 2}$ supersymmetric 
$U(1)$ gauge theory $Z_{NA}$ and its dual $Z_{NAD}$ are equivalent:
\[
Z_{NA}=Z_{NAD}.
\]
Therefore, under the strong-weak duality non--anticommutative $ N={1\over 
2}$ supersymmetric $ U(1)$ gauge theory is invariant. 

Loop corrections can 
be taken into account in terms of two different procedures. One of them is to 
calculate 
loop contributions to parent action and deduce the resulting theories from 
the loop corrected parent action. The other is to take into consideration 
loop corrections to  $N={1\over 2}$ supersymmetric $U(1)$
gauge theory \cite{laj} and 
then trying to formulate its dual. In the latter formulation it seems that 
our results survive with one loop corrected $C_{\m\n}$.


\begin{thebibliography}{99}

\bibitem{ber} N. Berkovits, {\it A new description of the superstrings,} 
\hepth{9604123}; {\it Super-Poincare 
covariant
quantization of the superstring}, JHEP 04 (2000) 018, \hepth{0001035}.

\bibitem{ov}H. Ooguri and C. Vafa, {\it The C-Deformation of Gluino and
Non-Planar Diagrams, Adv.Theor.Math.Phys.} {\bf 7} (2003) 53,
\hepth{0302109}.

\bibitem{sei} N. Seiberg, {\it Noncommutative superspace,
$N=1/2$
supersymmetry, field theory and string theory}, JHEP 06 (2003) 010,
\hepth{0305248}.

\bibitem{nie} J. de Boer, P.A. Grassi , P. van Nieuwenhuizen, {\it
Non-commutative superspace from string theory, Phys. Lett.} {\bf B}
{\bf 574} (2003) 98, \hepth{0302078}.

\bibitem{mof} J. W. Moffat, {\it Noncommutative and non-anticommutative 
quantum field theory, Phys. Lett.} {\bf B 506} (2001) 193, 
\hepth{0011035}.

\bibitem{klem} D. Klemm, S. Penati, L. Tamassia, {\it 
Non(anti)commutative superspace, Class.Quant.Grav.} {\bf 20} 
(2003) 2905, \hepth{0104190}.

\bibitem{lerda} M. Billo, M. Frau, I. Pesando, A. Lerda, {\it $N=1/2$ 
gauge theory and its instanton moduli space from open strings in R-R 
background,} JHEP 05 (2004) 023, \hepth{0402160}.

\bibitem{wb}J. Wess and J. Bagger,
{\em Supersymmetry and supergravity,} Princeton University Press,
Princeton, 1992.

\bibitem{kay}K. \"{U}lker, {\it $N=1$ SYM Action and BRST Cohomology,
Mod. Phys. Lett.} {\bf A17} (2002) 739,   \hepth{0108062};
{\it On the cohomological structure of supersymmetric lagrangians with and
without auxiliary fields, Mod. Phys. Let} {\bf A16} (2001) 881, 
\hepth{0011103}.

\bibitem{bus}T.H. Buscher,
{\it A symmetry of the string background field equations,
Phys. Lett.} $\mbf{B}$ $\mbf{194}$ (1987) 59;
{\it Path-integral derivation of
quantum duality in nonlinear sigma models, Phys. Lett. }
$\mbf{B}$ $\mbf{201}$ (1988) 466.

\bibitem{grs} O.J. Ganor, G. Rajesh and S. Sethi,
{\em Duality and noncommutative gauge theory,}
{\em Phys. Rev.} {\bf D 62} (2000) 125008, \hepth{0005046}.

\bibitem{okb}\"O.F. Dayi, K. \"{U}lker,
and  B. Yap{\i}\c{s}kan, {\it Duals of noncommutative supersymmetric $U(1)$
gauge theory}, JHEP 0310 (2003) 010, \hepth{0309073}.

\bibitem{bo} \"O.F. Dayi and  B. Yap{\i}\c{s}kan,
{\it Equivalence of partition
functions for  noncommutative $U(1)$ gauge theory and its dual in phase
space}, JHEP 0411 (2004) 064,  \hepth{0407269}.


\bibitem{sw-p}N. Seiberg and E. Witten,
{\em Electric--magnetic duality, monopole condensation,
and confinement in $N=2$ supersymmetric Yang--Mills
theory,} {\em  Nucl. Phys.} {\bf B 426}
(1994) 19, \hepth{9407087}.

\bibitem{fra}E. S. Fradkin, {\it Hamiltonian formalism in covariant gauge
and the measure in quantum gravity}
in {\it New Developments in Relativistic
Quantum Field Theory,} Proceedings of the Xth Winter School
of Theoretical Physics, Karpacz, Poland, 1973 [{\it Acta Univ.
Wratislav.} no. 207, Poland (1973)].

\bibitem{sen} P. Senjanovic, {\it Path integral quantization
of field theories with second class constraints,
Annals of Physics} {\bf 100} (1976) 227.

\bibitem{laj} O. Lunin, S. Rey, {\it Renormalizability of 
non(anti)commutative gauge theories with $N=1/2$ supersymmetry}, JHEP 09 
(2003) 045, \hepth{0307275};\\
 M. Alishahiha, A. Ghodsi, N. Sadooghi, {\it 
One-Loop Perturbative Corrections to non(anti)commutativity Parameter of 
$N=1/2$ 
Supersymmetric U(N) Gauge Theory, Nucl.Phys.} {\bf B 691} (2004) 111-128, 
\hepth{0309037};\\
 I. Jack, D.R.T. Jones, L.A. Worthy, {\it One-loop 
renormalisation of general N=1/2 supersymmetric gauge theory, Phys.Lett.B} 
{\bf 611} (2005) 199, \hepth{0412009}.

\end{thebibliography}
\end{document}